\documentclass[aps,prb,twocolumn,superscriptaddress,floatfix,longbibliography]{revtex4-2}
\usepackage{amsmath,amssymb,amsthm}
\usepackage{physics}
\usepackage{amsfonts}
\usepackage{mathrsfs}
\usepackage{graphicx}
\usepackage{tabularx}
\usepackage{enumerate}
\usepackage{dcolumn}
\usepackage{bm}
\usepackage{xcolor}
\usepackage[normalem]{ulem}
\usepackage[colorlinks,linkcolor=blue,citecolor=blue,urlcolor=blue]{hyperref}

\newcommand{\ii}{\mathrm{i}}

\newcommand{\Ree}{\operatorname{Re}}
\newcommand{\Imm}{\operatorname{Im}}

\begin{document}
	\title{Lee--Yang zeros of modulated XY spin chains with Dzyaloshinskii--Moriya interaction: zero-contour topology and quantum phase-diagram reconstruction}
	
	\author{Hong Jiang}
	\affiliation{Center of Materials Science and Optoelectronics Engineering, College of Materials Science and Opto-Electronic Technology, University of Chinese Academy of Sciences, Beijing 100049, China}
	
	\author{Xiang-Ping Jiang}
	\email{2015iopjxp@gmail.com}
	\affiliation{School of Physics, Hangzhou Normal University, Hangzhou, Zhejiang 311121, China}

	\author{Yan-Chao Li}
	\email{ycli@ucas.ac.cn}
	\affiliation{Center of Materials Science and Optoelectronics Engineering, College of Materials Science and Opto-Electronic Technology, University of Chinese Academy of Sciences, Beijing 100049, China}

	\begin{abstract}
		We investigate the Lee--Yang zeros (LYZ) of inhomogeneous anisotropic XY spin chains with Dzyaloshinskii--Moriya (DM) interactions in the complex transverse-field plane, focusing on their fundamental connection to quantum phase transitions. We systematically study uniform chains, period-2 and period-3 modulated chains, and Fibonacci quasiperiodic chains of lengths 5 and 8. As the DM coupling strength $D$ increases, the LYZ exhibit qualitatively distinct topological evolutions on the complex plane: the complex zeros of the uniform chain collapse toward the real axis; periodic chains feature either bifurcation of closed zero contours before all zeros become real or a single re-emergence of complex zeros; quasiperiodic chains exhibit repeated annihilation and revival of complex zeros. Analytical derivations demonstrate that this diverse behavior originates from DM-induced shifts of folded bands and the modulation of zero positions by anisotropic pairing at particle--hole band crossings. Our results establish a direct correspondence between LYZ topology and band deformation: isolated contact points of zeros with the real axis correspond to discrete quantum critical fields, while continuous real-zero intervals directly identify gapless chiral phases. Accordingly, beyond locating phase boundaries, LYZ can distinguish characteristic phases and serve as an intuitive probe for band folding and phase diagram restructuring.
	\end{abstract}
	
	\maketitle
	
	\section{Introduction}
	\label{sec:introduction}
	
	Lee and Yang demonstrated that classical phase transitions can be fully characterized by partition-function zeros obtained by analytically continuing the magnetic field into the complex plane \cite{YangLee1952,LeeYang1952,ItzyksonPearsonZuber1983,SuzukiFisher1971,BaakeGrimmPisani1995,BarataGoldbaum2001,BenaDrozLipowski2005}. At nonzero temperature, the zeros of a finite system do not lie on the real-field axis, but they accumulate and approach it in the thermodynamic limit as the system size increases. Their contact with the real axis gives the transition point, while their overall distribution describes how the singularity develops. This idea was later extended to quantum phase transitions. These transitions are driven by quantum control parameters such as the transverse field and coupling strength. When these parameters are analytically continued into the complex plane, the real-axis contacts of the Lee--Yang zeros (LYZ) likewise give the quantum critical points \cite{TongLiu2006,ZhongLiuTong2007,Kist2021}. Recent work has also connected complex partition-function zeros with quantum Fisher information, coherence, entanglement, fidelity, and quantum criticality in the complex inverse-temperature plane \cite{TaoEtAl2022,TaoEtAl2023,LiuLvYangZou2023,LiuLvMengTanZhaoZou2024,LiuYinChen2024,Li2025,GuSun2026,ZhangMaoHuZhaoSunYou2025,LvEtAl2026}. LYZ can now be calculated in interacting many-body systems and measured on quantum platforms using probe spins, trapped ions, and open-system dynamics \cite{WeiLiu2012,Peng2015,Francis2021,MatsumotoNakagawaUeda2022,VecseiLadoFlindt2022,VecseiFlindtLado2023,GaoEtAl2024,LanEtAl2024,ChatterjeeEtAl2024,VecseiLadoFlindt2025,WadaKitazawaKanaya2025}. Rydberg blockade models and atomic arrays provide another route to their measurement \cite{LiYang2023,ShenEtAl2023}.
	
	The spin-$1/2$ XY chain is a standard exactly solvable model for studying quantum phase transitions. The Jordan--Wigner transformation maps it to free fermions, allowing its excitation spectrum, correlations, and critical fields to be obtained directly \cite{LiebSchultzMattis1961,BarouchMcCoy1971}. When the exchange couplings follow a periodic or Fibonacci sequence, the modulation divides the original spectrum into several subbands and produces multiple critical fields \cite{SatijaDoria1988,Luck1993,Hermisson2000,TongZhong2001,TongZhong2002,deLima2007,TimoninChitov2021,ChitovGadgeTimonin2022,CaoFuLiuZhongTong2024}. Tong and Liu showed that these changes appear directly in the LYZ: one closed curve can split into several curves, and their intersections with the real axis agree with the critical fields of the corresponding periodic and Fibonacci XY chains \cite{TongLiu2006}. LYZ can therefore be used directly to study quantum phase transitions in exchange-modulated XY chains.
	
	The Dzyaloshinskii--Moriya (DM) interaction is an antisymmetric exchange interaction arising from spin--orbit coupling and broken spatial inversion symmetry \cite{Dzyaloshinsky1958,Moriya1960}. In the uniform XY chain, it makes the fermion dispersion asymmetric in momentum and, when sufficiently strong, produces a gapless chiral phase distinct from the gapped ordered and paramagnetic phases \cite{Liu2011,DerzhkoVerkholyakKrokhmalskiiButtner2006,JafariKargarianLangariSiahatgar2008,KadarZimboras2010,Zhong2013,YiDingRenWangYou2018,MahdavifarEtAl2024}. The effects of the DM interaction have also been studied in several modulated spin models \cite{WangYanYi2010,RoyEtAl2019,CornagliaEtAl2024}. Periodic and Fibonacci modulations reshape the excitation spectrum into multiple subbands. When the DM interaction is combined with these modulations, the phase structure may differ substantially from that of the uniform XY chain, and LYZ contours evolve in qualitatively different fashions. However, the quantum phase diagrams of such modulated systems, the corresponding LYZ topology, and their interconnection remain largely unexplored systematically.
	
	In this work, we derive the equations that determine the LYZ of inhomogeneous XY spin chains with the DM interaction and use the resulting zeros to characterize their quantum phase transitions. We take the uniform XY chain as a reference and then study the period-two and period-three XY chains, together with the $F_l=5$ and $F_l=8$ Fibonacci XY chains. We first determine the quantum phase boundaries from the real-field Bloch spectrum and use long-range spin correlations and vector chirality to distinguish the antiferromagnetic, paramagnetic, and gapless chiral phases. The LYZ are then obtained by analytically continuing the transverse field into the complex plane and solving the Bogoliubov--de Gennes (BdG) field equation. By comparing the phase boundaries obtained from the real-field Bloch spectrum with the real-axis intersections of the LYZ and the endpoints of the real-zero intervals, we show that the quantum phase boundaries in these systems can be identified clearly from the LYZ. The analytic field equations further show that the DM interaction moves the folded bands while anisotropy couples nearby branches, explaining why the LYZ contours split and reach the real axis and why complex curves reappear in the period-three and Fibonacci XY chains at larger $D$.
	
	The rest of the paper is organized as follows. Sec.~\ref{sec:model} introduces the model and calculation method. Secs.~\ref{sec:uniform} and \ref{sec:periodic} discuss the uniform and periodic XY chains, respectively. Sec.~\ref{sec:fibonacci} studies the Fibonacci XY chains, and Sec.~\ref{sec:conclusions} summarizes the main results. The detailed algebraic derivations for the short-period XY chains are given in the appendices.
	
	\section{Model and method}
	\label{sec:model}
	
	We consider inhomogeneous spin-$1/2$ XY chains with the DM interaction in a transverse field, described by the following Hamiltonian
	\begin{align}
		H=\frac12\sum_{j=1}^{N}\bigg\{&
		\lambda_j\left[
		\frac{1+\gamma}{2}\sigma_j^x\sigma_{j+1}^x+
		\frac{1-\gamma}{2}\sigma_j^y\sigma_{j+1}^y\right]\nonumber\\
		&+D\left(\sigma_j^x\sigma_{j+1}^y-
		\sigma_j^y\sigma_{j+1}^x\right)
		+h\sigma_j^z\bigg\}.
		\label{eq:spinH}
	\end{align}
	The nearest-neighbor interaction $\lambda_j$ is either $1$ or $\alpha$, $\gamma$ is the XY anisotropy, and $D$ is the DM coupling strength.  In this work, we study the uniform cell, the period-two cell $(1,\alpha)$, and the period-three cell $(1,\alpha,\alpha)$.  We also study two Fibonacci XY chains.  Let $A$ denote a bond with $\lambda_j=\alpha$ and $B$ a bond with $\lambda_j=1$.  The recursion
	\begin{equation}
		S_0=B,\qquad S_1=A,\qquad S_{l+1}=S_lS_{l-1}
		\label{eq:fib_recursion}
	\end{equation}
	gives
	\begin{equation}
		S_4=ABAAB,\qquad S_5=ABAABABA .
		\label{eq:fib_cells}
	\end{equation}
	For the Bloch calculation, $S_4$ and $S_5$ are repeated along the chain as five-bond and eight-bond unit cells.  We refer to them as the $F_l=5$ and $F_l=8$ Fibonacci XY chains below. Throughout this work, without loss of generality, we set the parameters $\gamma=\alpha=0.5$.
	
	After a Jordan--Wigner transformation~\cite{LiebSchultzMattis1961}, Eq.~\eqref{eq:spinH} becomes
	\begin{equation}
		H=\frac{Nh}{2}
		+\sum_{i,j=1}^{N}
		\left[
		c_i^\dagger A_{ij}c_j
		+\frac12\left(
		c_i^\dagger B_{ij}c_j^\dagger+\mathrm{H.c.}
		\right)
		\right],
		\label{eq:fermion_hamiltonian}
	\end{equation}
	where $c_j$ annihilates a spinless fermion at site $j$.  The nonzero matrix elements are $A_{jj}=-h$, $A_{j,j+1}=\lambda_j/2+\ii D$, $A_{j+1,j}=\lambda_j/2-\ii D$, $B_{j,j+1}=\gamma\lambda_j/2$, and $B_{j+1,j}=-\gamma\lambda_j/2$.  For real $h$, $A=A^\dagger$ and $B=-B^T$.  The exchange and DM interactions give the real and imaginary parts of the hopping, while the anisotropy gives the pairing term \cite{LiebSchultzMattis1961,BarouchMcCoy1971,TongLiu2006}.
	
	For a unit cell containing $p$ sites, the total chain length is $N=pN_c$, where $N_c$ is the number of unit cells.  The models considered below have $p=1$, $2$, $3$, $5$, or $8$.  Writing
	\begin{equation}
		\begin{aligned}
			c_{\ell,a}
			&=\frac{1}{\sqrt{N_c}}\sum_q e^{\ii q\ell}c_{q,a},\\
			\Psi_q
			&=(c_{q,1},\ldots,c_{q,p},
			c_{-q,1}^\dagger,\ldots,c_{-q,p}^\dagger)^T,
		\end{aligned}
		\label{eq:bloch_nambu_basis}
	\end{equation}
	where $\ell$ labels the unit cell and $a$ labels a site inside it, the $2p\times2p$ Bloch BdG matrix is
	\begin{equation}
		\mathcal H_p(q,h)=
		\begin{pmatrix}
			\mathbf A_p(q,h) & \mathbf B_p(q)\\
			-\mathbf B_p^*(-q) & -\mathbf A_p^T(-q,h)
		\end{pmatrix}.
		\label{eq:bloch_bdg}
	\end{equation}
	Here $\mathbf A_p(q,h)$ and $\mathbf B_p(q)$ are the Bloch forms of $A$ and $B$ defined above.  The bond joining neighboring unit cells carries the factors $e^{\ii q}$ and $e^{-\ii q}$.  Since the transverse field enters only through $A_{jj}=-h$,
	\begin{equation}
		\mathcal H_p(q,h)
		=
		\mathcal H_p(q,0)
		+h
		\begin{pmatrix}
			-I_p&0\\
			0&I_p
		\end{pmatrix}.
		\label{eq:bdg_pencil}
	\end{equation}
	
	For a real field, the quasiparticle gap is
	\begin{equation}
		\Delta(h,D)=\min_{q,n}\left|E_n(q,h,D)\right|,
		\label{eq:physical_gap}
	\end{equation}
	where $E_n$ are the eigenvalues of $\mathcal H_p$. To characterize the real-field phases, we calculate the long-distance correlation
	\begin{equation}
		C_x(r)=\frac{1}{N}\sum_j
		\langle\sigma_j^x\sigma_{j+r}^x\rangle .
		\label{eq:correlation}
	\end{equation}
	The vector chirality is
	\begin{equation}
		\kappa=\frac{1}{N}\sum_j
		\left\langle\sigma_j^x\sigma_{j+1}^y-
		\sigma_j^y\sigma_{j+1}^x\right\rangle .
		\label{eq:chirality}
	\end{equation}
	For all models considered below, the phase boundaries shown in panel (a) are obtained from the gap-closing condition in Eq.~\eqref{eq:physical_gap}, while the phase regions are identified using the long-distance correlation in Eq.~\eqref{eq:correlation} and the vector chirality in Eq.~\eqref{eq:chirality}. With the sign convention of Eq.~\eqref{eq:spinH}, these quantities distinguish three phases.  The gapped antiferromagnetic (AFM) phase has a finite gap and a nonzero long-distance $x$-spin correlation, the gapped paramagnetic (PM) phase has a finite gap but no long-range order, and the gapless chiral phase has a vanishing gap and finite vector chirality \cite{YiDingRenWangYou2018,MahdavifarEtAl2024}.  Since the DM interaction tilts neighboring spins, a modulated gapped phase may also have finite chirality.  We therefore use the gap and the long-distance correlation together to distinguish it from the gapless chiral phase.
	
	After the Bogoliubov diagonalization, the partition function is
	\begin{equation}
		\begin{aligned}
			Z(h,T)
			&=\operatorname{Tr}\left(e^{-\beta H}\right) =\prod_{q,n} 2\cosh\left[\frac{\beta E_n(q,h)}{2}\right],
		\end{aligned}
	\end{equation}
	where $\beta=(k_B T)^{-1}$ and the product is taken over the independent quasiparticle modes. The zeros of $Z$ satisfy
	\begin{equation}
		E_n(q,h)=i(2m+1)\pi k_B T,
		\qquad m\in\mathbb{Z}.
	\end{equation}
	Defining $t=(2m+1)\pi k_B T$, the field zeros follow from
	\begin{equation}
		\left[\ii t I_{2p}-\mathcal H_p(q,0)\right]v
		=h
		\begin{pmatrix}
			-I_p&0\\
			0&I_p
		\end{pmatrix}v .
		\label{eq:field_pencil}
	\end{equation}
	We focus on $t=0$, where Eq.~\eqref{eq:field_pencil} gives all complex values of $h$ at which a quasiparticle energy vanishes.  Its real solutions are the gap closings of Eq.~\eqref{eq:physical_gap}.  In the following sections, these LYZ are plotted in blue and compared with the isolated critical fields and gapless field intervals obtained independently from the real-field Bloch spectrum, shown as red points and red segments, respectively.  Their agreement on the real axis directly shows how the LYZ mark the physical phase boundaries.
	
	\section{Uniform XY chain}
	\label{sec:uniform}
		
	\begin{figure}[!t]
		\centering
		\includegraphics[width=\columnwidth]{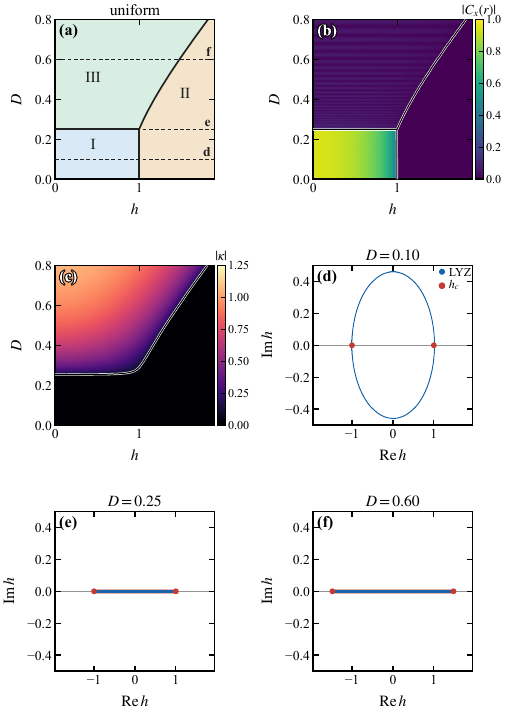}
		\caption{\label{fig:uniform} Uniform XY chain at $\gamma=0.5$.  (a) Real-field phase diagram, where I, II, and III denote the AFM, PM, and gapless chiral phases. (b) Long-distance correlation $|C_x(r)|$. (c) Vector chirality $|\kappa|$. (d)--(f) LYZ at $D=0.10$, $0.25$, and $0.60$, corresponding to the values of $D$ marked by the dashed lines in panel (a).  The heat maps use $N_c=512$, while the LYZ use $N_c=32768$, with $N=N_c$.  Blue points are LYZ, red points mark $h_c$, and a red segment marks a gapless interval.}
	\end{figure}
	
	The uniform XY chain gives the simplest view of the effect of the DM coupling. In a convenient Nambu basis, its two quasiparticle energies are
	\begin{equation}
		E_\pm(k)=-2D\sin k
		\pm\sqrt{(h-\cos k)^2+\gamma^2\sin^2 k}.
		\label{eq:uniform_energy}
	\end{equation}
	Setting either energy to zero gives the complex-field curves
	\begin{equation}
		h_\pm(k)=\cos k\pm
		\sqrt{4D^2-\gamma^2}\,\sin k .
		\label{eq:uniform_zeros}
	\end{equation}
	For $D<\frac{\gamma}{2}$, the square root is imaginary and the zeros form an ellipse. Its two real intersections remain at $h=\pm1$.  At $D=\frac{\gamma}{2}$, the ellipse is compressed onto the segment $-1\le h\le1$.  For $D>\frac{\gamma}{2}$, all zeros are real and fill
	\begin{equation}
		|h|\le h_c,\qquad
		h_c=\sqrt{1+4D^2-\gamma^2}.
		\label{eq:uniform_hc}
	\end{equation}
	Thus the weak-DM transition occurs at $|h|=1$, while the strong-DM gapless phase ends at the field in Eq.~\eqref{eq:uniform_hc}.  The three boundaries are the vertical line $h=1$, the horizontal line $D=\gamma/2$, and the curve $h=h_c$.  They give the AFM, PM, and gapless chiral regions of the phase diagram of the uniform XY chain \cite{Zhong2013,YiDingRenWangYou2018,MahdavifarEtAl2024}.
	
	Figure~\ref{fig:uniform}(a) shows the phase diagram obtained from the gap closings of the spectrum in Eq.~\eqref{eq:uniform_energy}. Panels (b) and (c) plot the long-distance correlation in Eq.~\eqref{eq:correlation} and the vector chirality in Eq.~\eqref{eq:chirality}, respectively.  The correlation is finite in region I, while the chirality becomes large in region III, confirming the phases obtained from the real spectrum.  At $D=0.10$, Fig.~\ref{fig:uniform}(d) shows that the zeros form an ellipse and touch the real axis at the AFM--PM critical fields. In Fig.~\ref{fig:uniform}(e), the ellipse becomes a real segment at $D=0.25=\gamma/2$. At $D=0.60$, Fig.~\ref{fig:uniform}(f) shows a longer segment whose endpoints coincide with the chiral--PM boundary given by Eq.~\eqref{eq:uniform_hc}. Real zeros inside the segment belong to the gapless phase, while its endpoints are the phase boundaries. The phase boundaries and LYZ in Fig.~\ref{fig:uniform} are in full agreement with the analytic results for the uniform XY chain with the DM interaction in Ref.~\cite{Zhong2013}.

	\section{Periodic XY chains}
	\label{sec:periodic}
	
	Exchange modulation folds the uniform-chain spectrum into several bands, producing more real-field crossings and LYZ curves. Tong and Liu showed that decreasing $\gamma$ splits the LYZ of periodic and Fibonacci-modulated XY chains into an increasing number of closed curves \cite{TongLiu2006}. Here $\gamma$ is fixed, and varying $D$ alone also moves the folded bands and produces a similar splitting. In the period-two XY chain, one closed curve splits into two before both become real. In the period-three XY chain, another pair of folded bands meets at larger $D$, causing a complex curve to reappear after all the zeros have become real.
	
	\subsection{Period-two XY chain}
	\label{sec:period2}
	
	The period-two XY chain has the cell $(1,\alpha)$.  Its hopping coefficients are
	\begin{equation}
		u_1=\frac12+\ii D,\qquad
		u_2=\frac{\alpha}{2}+\ii D.
		\label{eq:p2_hoppings}
	\end{equation}
	In the Nambu basis of Eq.~\eqref{eq:bloch_nambu_basis}, the $4\times4$ BdG matrix in Eq.~\eqref{eq:bloch_bdg} is formed from
	\begin{equation}
		\mathbf A_2(q,h)=
		\begin{pmatrix}
			-h&u_1+u_2^*e^{-\ii q}\\
			u_1^*+u_2e^{\ii q}&-h
		\end{pmatrix},
		\label{eq:p2_normal_block}
	\end{equation}
	and
	\begin{equation}
		\mathbf B_2(q)=
		\frac{\gamma}{2}
		\begin{pmatrix}
			0&1-\alpha e^{-\ii q}\\
			-1+\alpha e^{\ii q}&0
		\end{pmatrix}.
		\label{eq:p2_pairing_block}
	\end{equation}
	Diagonalizing $\mathcal H_2(q,h)$ gives four quasiparticle bands.  Their exact characteristic equation is
	\begin{align}
		{}&E^4-
		\Bigg\{
		2h^2+\frac12\Big[
		(1+\alpha^2)(1+\gamma^2)
		+2\alpha(1-\gamma^2)\cos q\nonumber\\
		&\hspace{20mm}
		+8D^2(1-\cos q)
		\Big]
		\Bigg\}E^2\nonumber\\
		&-4(1+\alpha)Dh\sin q\,E
		+\det\mathcal H_2(q,h)=0 .
		\label{eq:p2_energy_polynomial}
	\end{align}
	For real $h$, the four roots give the spectrum used in Eq.~\eqref{eq:physical_gap}.  Their explicit analytic form is derived in Appendix~\ref{app:p2_bands}. 
	
	At $t=0$, setting $E=0$ gives the field equation $P_2(h,q)\equiv\det\mathcal H_2(q,h)=0$.  We define
	\begin{align}
		R_2&=(1+\alpha)^2-\gamma^2(1-\alpha)^2,
		\label{eq:p2_R}\\
		S_2&=\gamma^2(1+\alpha)^2-(1-\alpha)^2.
		\label{eq:p2_S}
	\end{align}
	The exact field equation then contains only even powers of $h$:
	\begin{equation}
		P_2(h,q)=
		\left[h^2-A_2(q)\right]^2-\Delta_2(q)=0,
		\label{eq:p2_polynomial}
	\end{equation}
	where
	\begin{align}
		A_2(q)&=\frac14\left[
		R_2\cos^2\frac q2+
		\left(16D^2-S_2\right)\sin^2\frac q2
		\right],
		\label{eq:p2_A}\\
		\Delta_2(q)&=\sin^2q
		\left(D^2R_2-\alpha^2\gamma^2\right).
		\label{eq:p2_discriminant}
	\end{align}
	Here $A_2(q)$ is the average of the two solutions for $h^2$.  The sign of $\Delta_2(q)$ determines whether they are real or complex.  This gives four field zeros for each unit-cell momentum $q$:
	\begin{equation}
		h(q)=\pm\sqrt{
			A_2(q)\pm\sqrt{\Delta_2(q)}}.
		\label{eq:p2_roots}
	\end{equation}
	The two signs are chosen independently.
	
	At $q=0$ and $q=\pi$, $\Delta_2(q)=0$.  The corresponding values of $h^2$ are
	\begin{equation}
		h_0^2=\frac{R_2}{4},\qquad
		h_\pi^2=4D^2-\frac{S_2}{4}.
		\label{eq:p2_contacts}
	\end{equation}
	When $S_2>0$, the pair of zeros at $q=\pi$ approaches the real axis as $D$ increases and reaches it at $h_\pi=0$.  This gives the threshold at which the closed zero curve splits,
	\begin{equation}
		D_{\rm split}^{(2)}
		=\frac{\sqrt{S_2}}{4}.
		\label{eq:p2_split}
	\end{equation}
	
	\begin{figure}[!t]
		\centering
		\includegraphics[width=\columnwidth]{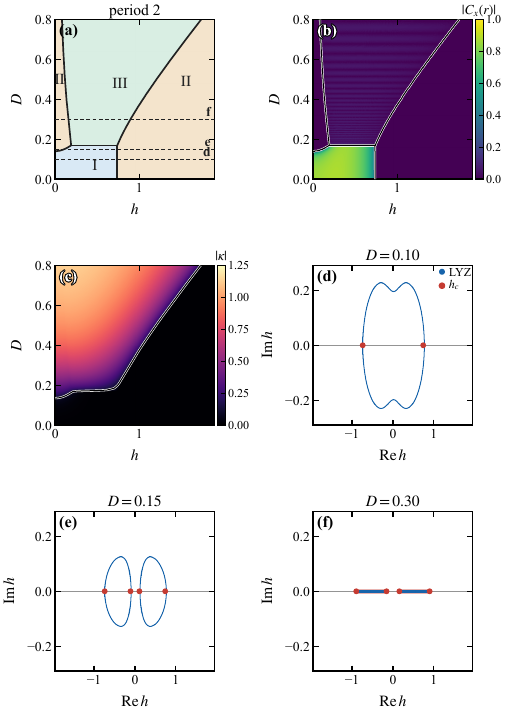}
		\caption{\label{fig:period2} Period-two XY chain with the cell $(1,\alpha)$ and $\gamma=\alpha=0.5$.  (a) Real-field phase diagram.  (b) Long-distance correlation $|C_x(r)|$.  (c) Vector chirality $|\kappa|$.  (d)--(f) LYZ at $D=0.10$, $0.15$, and $0.30$, corresponding to the values of $D$ marked by the dashed lines in panel (a).  The heat maps use $N_c=512$ unit-cell Bloch momenta and $N=2N_c=1024$ sites, while the LYZ use $N_c=32768$ and $N=2N_c=65536$.  Blue points are the LYZ, red points mark the critical fields, and red segments mark the gapless intervals.}
	\end{figure}
	
	The sign of $\Delta_2(q)$ determines when all the zeros become real.  Setting $D^2R_2-\alpha^2\gamma^2=0$ gives
	\begin{equation}
		D_{\rm real}^{(2)}
		=\frac{\alpha\gamma}{\sqrt{R_2}}.
		\label{eq:p2_real}
	\end{equation}
	At this threshold, the last pair of complex zeros reaches the real axis.  It is also easy to see that the two thresholds satisfy
	\begin{equation}
		\left(D_{\rm real}^{(2)}\right)^2
		-\left(D_{\rm split}^{(2)}\right)^2
		=
		\frac{(1-\alpha^2)^2(1-\gamma^2)^2}{16R_2}
		\geq0.
		\label{eq:p2_threshold_order}
	\end{equation}
	Therefore, when the splitting threshold exists, one zero curve first splits into two curves, and all the zeros then move onto the real axis.  For $\alpha=1$, $R_2=4$ and $S_2=4\gamma^2$.  The period-two cell then simply folds the bands of the uniform XY chain.  The right-hand side of Eq.~\eqref{eq:p2_threshold_order} is zero, and both thresholds coincide at $\gamma/2$.  The region with two closed curves therefore disappears, and the equation for the zeros reduces to that of the uniform XY chain.
	
	For $\alpha=\gamma=0.5$, $R_2=35/16$ and $S_2=5/16$.  The two thresholds are $D_{\rm split}^{(2)}=\sqrt{5}/16\simeq0.1398$ and $D_{\rm real}^{(2)}=1/\sqrt{35}\simeq0.1690$.  For $D<D_{\rm split}^{(2)}$, the zeros form one closed curve.  Between the two thresholds, they form two closed curves.  For $D>D_{\rm real}^{(2)}$, all the zeros lie on the real axis.
	
	For real $h$, the same zero-energy condition gives the physical phase boundaries.  In the strong-DM region, the gapless interval on the positive real axis is $h_L\leq h\leq h_R$, where
	\begin{equation}
		h_{L,R}=\frac12\left[
		\sqrt{1+4D^2-\gamma^2}
		\mp\sqrt{4D^2+\alpha^2(1-\gamma^2)}
		\right].
		\label{eq:p2_interval}
	\end{equation}
	The corresponding interval on the negative real axis follows from symmetry.  All the real zeros therefore lie in $[-h_R,-h_L]\cup[h_L,h_R]$.  The inner and outer endpoints give the boundaries between the gapless chiral phase and the neighboring gapped phases.
	
	Figure~\ref{fig:period2}(a) shows the phase diagram obtained from the gap closings of the four energy bands determined by Eq.~\eqref{eq:p2_energy_polynomial}. The phase boundaries agree with those obtained from the ground-state geometric phase for the corresponding period-two XY chain in Ref.~\cite{WangYanYi2010}. In Fig.~\ref{fig:period2}(b), the long-distance correlation is finite in the antiferromagnetic region.  In Fig.~\ref{fig:period2}(c), the DM interaction keeps the vector chirality finite over a wider range of $h$ and $D$.  Together with the energy gap, these two quantities distinguish the three phases.  At $D=0.10$, the zeros in Fig.~\ref{fig:period2}(d) form one closed curve.  Exchange modulation changes its shape, and the curve intersects the real axis at several critical fields.  At $D=0.15$, the curve in Fig.~\ref{fig:period2}(e) has split into two, and both curves meet the real axis at the critical fields marked in red.  At $D=0.30$, all the zeros in Fig.~\ref{fig:period2}(f) lie on two real intervals.  Their inner and outer endpoints agree with Eq.~\eqref{eq:p2_interval}.  The factor $D^2R_2-\alpha^2\gamma^2$ in $\Delta_2(q)$ changes sign only once as $D$ increases.  Therefore, after all the zeros of the period-two XY chain have moved onto the real axis, they do not form a complex curve again at larger $D$.  The period-three XY chain discussed next behaves differently.

	\subsection{Period-three XY chain}
	\label{sec:period3}
	
	\begin{figure*}[t]
		\centering
		\includegraphics[width=0.98\textwidth]{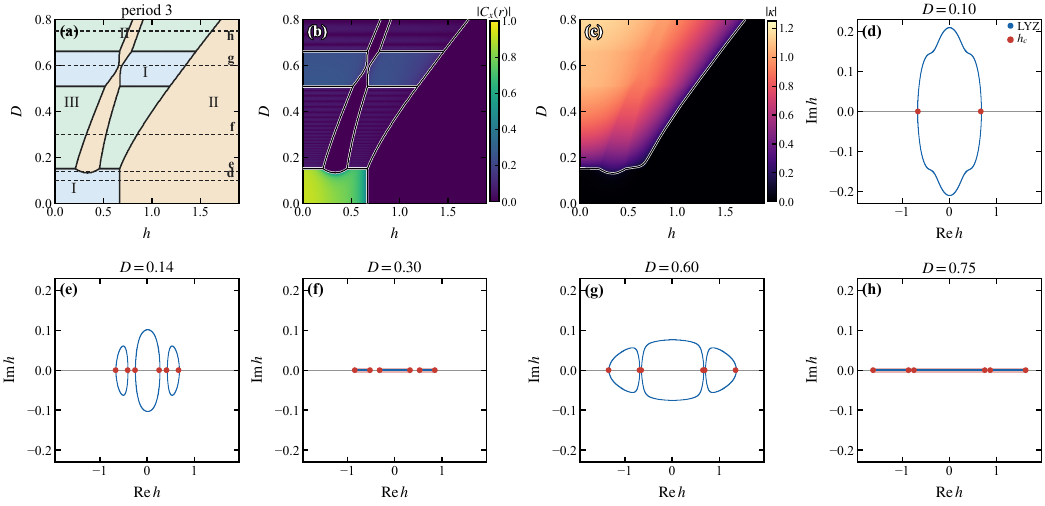}
		\caption{\label{fig:period3} Period-three XY chain with $(1,\alpha,\alpha)$ and $\gamma=\alpha=0.5$.  (a) Real-field phase diagram.  (b) $|C_x(r)|$.  (c) $|\kappa|$.  (d)--(h) LYZ at $D=0.10$, $0.14$, $0.30$, $0.60$, and $0.75$, corresponding to the values of $D$ marked by the dashed lines in panel (a).  The heat maps use $N_c=512$ and $N=1536$, while the LYZ use $N_c=32768$ and $N=98304$.  A complex curve reappears in panel (g) and becomes real again in panel (h).}
	\end{figure*}
	
	The period-three XY chain has the cell $(1,\alpha,\alpha)$.  Its hopping coefficients are
	\begin{equation}
		u_1=\frac12+\ii D,\qquad
		u_2=u_3=\frac{\alpha}{2}+\ii D.
		\label{eq:p3_hoppings}
	\end{equation}
	In the Nambu basis of Eq.~\eqref{eq:bloch_nambu_basis}, the $6\times6$ BdG matrix is formed from
	\begin{equation}
		\mathbf A_3(q,h)=
		\begin{pmatrix}
			-h&u_1&u_3^*e^{-\ii q}\\
			u_1^*&-h&u_2\\
			u_3e^{\ii q}&u_2^*&-h
		\end{pmatrix},
		\label{eq:p3_normal_block}
	\end{equation}
	and
	\begin{equation}
		\mathbf B_3(q)=
		\frac{\gamma}{2}
		\begin{pmatrix}
			0&1&-\alpha e^{-\ii q}\\
			-1&0&\alpha\\
			\alpha e^{\ii q}&-\alpha&0
		\end{pmatrix}.
		\label{eq:p3_pairing_block}
	\end{equation}
	The six energy bands satisfy
	\begin{equation}
		\det\!\left[E I_6-\mathcal H_3(q,h)\right]=0.
		\label{eq:p3_energy_polynomial}
	\end{equation}
	At a general $q$, this sixth-degree equation has no general solution in radicals, so we obtain the bands by numerically diagonalizing $\mathcal H_3(q,h)$. For real $h$, the resulting excitation gap determines the phase boundaries shown in Fig.~\ref{fig:period3}(a).
	
	At $t=0$, setting $E=0$ reduces Eq.~\eqref{eq:p3_energy_polynomial} to the field equation for $h$,
	\begin{equation}
		P_3(h,q)\equiv\det\mathcal H_3(q,h)=0.
		\label{eq:p3_field_equation}
	\end{equation}
	For each momentum $q$, we solve this sixth-degree polynomial numerically for its six roots in $h$ using Eq.~\eqref{eq:field_pencil}. At $q=0$ and $\pi$, the polynomial factorizes into the square of a cubic, while at $q=\pi/2$ it becomes a cubic polynomial in $h^2$. These simpler forms give the real-axis contacts and explain why a complex zero curve reappears at larger $D$.
	
	At $q=0$ and $q=\pi$, the sixth-degree determinant is the square of a cubic polynomial.  Defining
	\begin{align}
		A_3={}&(2\alpha^2+1)(\gamma^2-1)-12D^2,
		\label{eq:p3_A}\\
		B_3={}&4D^2(2\alpha+1)-\alpha^2(3\gamma^2+1).
		\label{eq:p3_B}
	\end{align}
	we obtain
	\begin{equation}
		\begin{aligned}
			P_3(h,0)
			&=-\frac{\left(4h^3+A_3h+B_3\right)^2}{16},\\
			P_3(h,\pi)
			&=-\frac{\left(4h^3+A_3h-B_3\right)^2}{16}.
		\end{aligned}
		\label{eq:p3_contact_cubic}
	\end{equation}
	The square accounts for the six field zeros.  The real roots of the first cubic give the contacts at $q=0$, and those of the second give the contacts at $q=\pi$.  The number of real roots changes when
	\begin{equation}
		A_3^3+27B_3^2=0.
		\label{eq:p3_contact_discriminant}
	\end{equation}
	For the parameters used in Fig.~\ref{fig:period3}, $\alpha=\gamma=0.5$, the positive solution of this condition gives
	\begin{equation}
		D_{\rm split}^{(3)}\simeq0.1325.
		\label{eq:p3_split_threshold}
	\end{equation}
	At this value, the single closed curve splits into three.
	These roots give the high-symmetry contacts in Fig.~\ref{fig:period3}(a).  Other endpoints occur at general momenta and are obtained from the zero of the full real-field Bloch spectrum.
	
	We next explain why complex zeros return at larger $D$. When $\gamma=0$, $\mathbf B_3(q)=0$ and the two diagonal
	blocks of $\mathcal H_3(q,h)$ decouple. Writing
	\begin{equation}
		\mathbf A_3(q,h)=K_3(q)-hI_3,
	\end{equation}
	the hopping matrix is
	\begin{equation}
		K_3(q)=
		\begin{pmatrix}
			0&u_1&u_3^*e^{-\ii q}\\
			u_1^*&0&u_2\\
			u_3e^{\ii q}&u_2^*&0
		\end{pmatrix}.
		\label{eq:p3_particle_block}
	\end{equation}
	With $\Phi_3=u_1u_2u_3$, the particle and hole field zeros satisfy
	\begin{align}
		\chi_{\rm p}(h,q)
		={}&h^3-\left(|u_1|^2+|u_2|^2+|u_3|^2\right)h\nonumber\\
		&
		-2\Ree\left(e^{\ii q}\Phi_3\right)=0,
		\label{eq:p3_particle_polynomial}\\
		\chi_{\rm h}(h,q)
		={}&h^3-\left(|u_1|^2+|u_2|^2+|u_3|^2\right)h\nonumber\\
		&
		-2\Ree\left(e^{-\ii q}\Phi_3\right)=0.
		\label{eq:p3_hole_polynomial}
	\end{align}
	Their difference is
	\begin{equation}
		\chi_{\rm p}-\chi_{\rm h}
		=4\sin q\,\Imm\Phi_3.
		\label{eq:p3_particle_hole_difference}
	\end{equation}
	At $q=\pi/2$, a particle root and a hole root meet at $h=0$ when $\Imm\Phi_3=0$.  Directly evaluating the product gives
	\begin{equation}
		\Imm\Phi_3
		=\frac{D}{4}\left(\alpha^2+2\alpha-4D^2\right).
		\label{eq:p3_im_phi}
	\end{equation}
	Besides $D=0$, this expression vanishes at
	\begin{equation}
		D_*=\frac12\sqrt{\alpha(\alpha+2)} .
		\label{eq:p3_Dstar}
	\end{equation}
	Thus, two folded particle and hole bands meet again at $D_*$.  For the period-two cell, the corresponding imaginary part is $D(1+\alpha)/2$ and has no second zero at $D>0$, so this additional meeting does not occur.
	
	For $\gamma\ne0$, the anisotropic pairing couples the two roots that meet at $D_*$.  Define
	\begin{align}
		F_\alpha(D,\gamma)={}&-8D^3
		+4(1-2\alpha)\gamma D^2\nonumber\\
		&+2\alpha\left[\alpha+2+(2-\alpha)\gamma^2\right]D
		+\alpha^2\gamma(\gamma^2+3).
		\label{eq:p3_F}
	\end{align}
	At $q=\pi/2$, the exact sixth-degree field equation becomes
	\begin{equation}
		h^2\left(h^2+\frac{A_3}{4}\right)^2
		+\frac{F_\alpha(D,\gamma)F_\alpha(-D,\gamma)}{16}=0.
		\label{eq:p3_resonance}
	\end{equation}
	The roots at $h=0$ merge when either $F_\alpha(D,\gamma)$ or $F_\alpha(-D,\gamma)$ vanishes.  For $\alpha=\gamma=0.5$, these conditions become
	\begin{equation}
		256D^3-92D-13=0,\qquad
		256D^3-92D+13=0 .
		\label{eq:p3_merger_cubics}
	\end{equation}
	For the same parameters, the three positive solutions give
	\begin{equation}
		\begin{aligned}
			D_{\rm real}^{(3)}&\simeq0.1509,\\
			D_-^{(3)}&\simeq0.5096,\qquad
			D_+^{(3)}\simeq0.6605.
		\end{aligned}
		\label{eq:p3_zero_thresholds}
	\end{equation}
	At $D_{\rm real}^{(3)}$, the $q=\pi/2$ roots meet at $h=0$, and the three low-$D$ complex curves become real. Complex zeros reappear for $D_-^{(3)}<D<D_+^{(3)}$. Equation~\eqref{eq:p3_Dstar} gives $D_*=0.5590$ for $\alpha=0.5$, so $D_-^{(3)}<D_*<D_+^{(3)}$.
	
	Figure~\ref{fig:period3}(a) shows the real-field phase diagram of the period-three XY chain. The long-distance correlation and vector chirality in Figs.~\ref{fig:period3}(b) and \ref{fig:period3}(c), together with the excitation gap, distinguish the three phases. Figs.~\ref{fig:period3}(d)--~\ref{fig:period3}(h) show how the LYZ change as $D$ increases. At $D=0.10<D_{\rm split}^{(3)}$, the zeros form one closed curve. The value $D=0.14$ in Fig.~\ref{fig:period3}(e) lies between $D_{\rm split}^{(3)}$ and $D_{\rm real}^{(3)}$, where the zeros form three smaller closed curves. At $D=0.30$, with $D_{\rm real}^{(3)}<D<D_-^{(3)}$, all the zeros lie on three real intervals. Complex curves reappear at $D=0.60$, which lies between $D_-^{(3)}$ and $D_+^{(3)}$, and all the zeros are real again at $D=0.75>D_+^{(3)}$. Throughout this sequence, the real-axis intersections and the endpoints of the real intervals coincide with the critical fields in Fig.~\ref{fig:period3}(a). Near $D_*$, the anisotropy moves the roots that meet around $q=\pi/2$ into the complex field plane; the corresponding expansion is given in Appendix~\ref{app:p3}.
	
	\section{Fibonacci XY chains}
	\label{sec:fibonacci}
	
	\begin{figure*}[t]
		\centering
		\includegraphics[width=\textwidth]{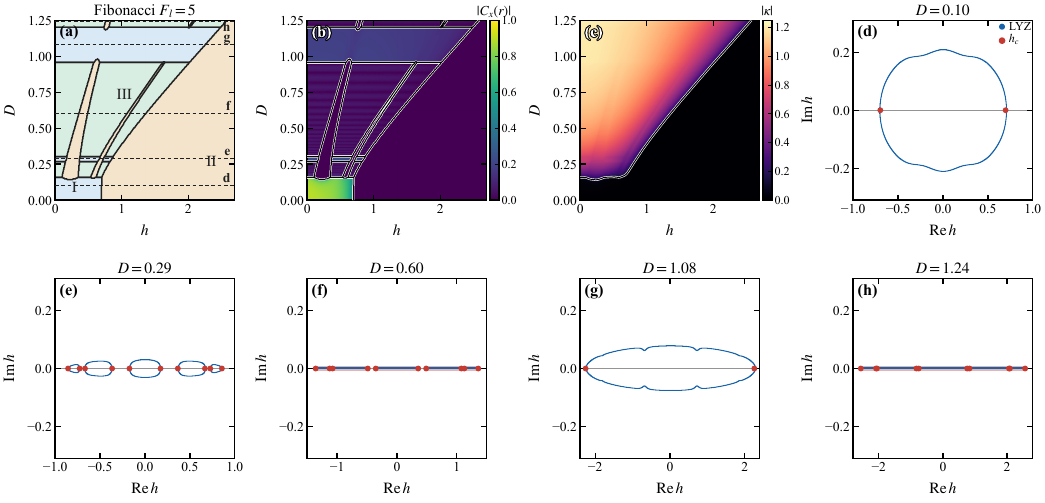}
		\caption{\label{fig:fib5} $F_l=5$ Fibonacci XY chain $ABAAB$ at $\gamma=\alpha=0.5$.  (a) Real-field phase diagram.  (b) Long-distance correlation $|C_x(r)|$.  (c) Vector chirality $|\kappa|$.  (d)--(h) LYZ at $D=0.10$, $0.29$, $0.60$, $1.08$, and $1.24$, corresponding to the values of $D$ marked by the dashed lines in panel (a).  The heat maps use $N_c=512$ and $N=2560$, while the LYZ use $N_c=32768$ and $N=163840$.}
	\end{figure*}
	
\begin{figure*}[t]
		\centering
		\includegraphics[width=\textwidth]{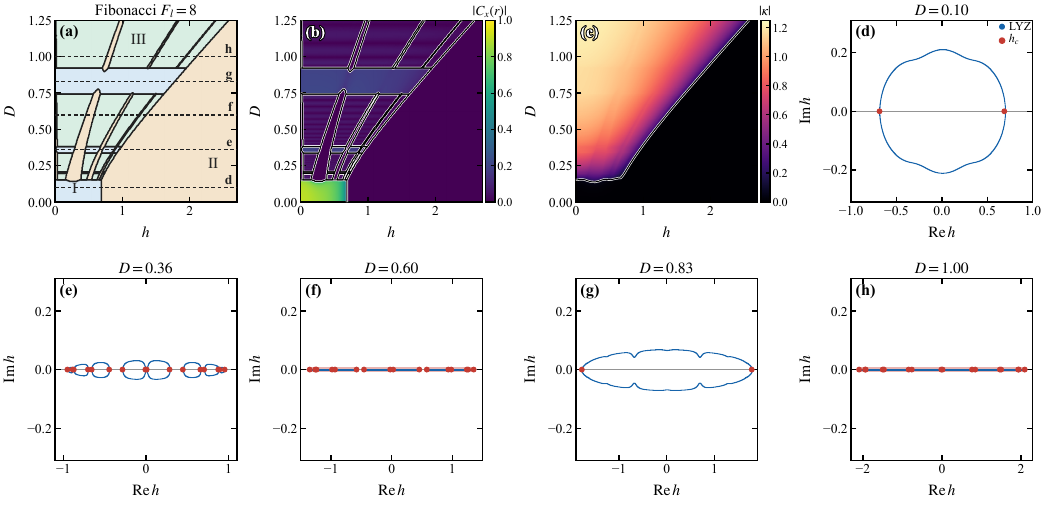}
		\caption{\label{fig:fib8} $F_l=8$ Fibonacci XY chain $ABAABABA$ at $\gamma=\alpha=0.5$.  (a) Real-field phase diagram.  (b) Long-distance correlation $|C_x(r)|$.  (c) Vector chirality $|\kappa|$.  (d)--(h) LYZ at $D=0.10$, $0.36$, $0.60$, $0.83$, and $1.00$, corresponding to the values of $D$ marked by the dashed lines in panel (a).  The heat maps use $N_c=512$ and $N=4096$, while the LYZ use $N_c=32768$ and $N=262144$.}
	\end{figure*}

	The Fibonacci cells are longer than the period-two and period-three cells discussed above. For the $F_l=5$ sequence $ABAAB$, the Bloch BdG matrix defined in Eq.~\eqref{eq:bloch_bdg} is $10\times10$; for the $F_l=8$ sequence $ABAABABA$, it is $16\times16$. Their energy bands satisfy
	\begin{equation}
		\begin{aligned}
			\det\!\left[E I_{10}-\mathcal H_5(q,h)\right]&=0,\\
			\det\!\left[E I_{16}-\mathcal H_8(q,h)\right]&=0.
		\end{aligned}
		\label{eq:fibonacci_energy_equations}
	\end{equation}
	At a general momentum $q$, these are tenth- and sixteenth-degree equations in $E$ and have no general solutions in radicals. We therefore obtain the bands by numerically diagonalizing $\mathcal H_5(q,h)$ and $\mathcal H_8(q,h)$. For real $h$, we obtain the phase boundaries from the excitation gap, as shown in Figs.~\ref{fig:fib5}(a) and \ref{fig:fib8}(a). The long-distance correlation and vector chirality further distinguish the antiferromagnetic, paramagnetic, and gapless chiral phases.
	
	At zero temperature, setting $E=0$ turns the same determinants into equations for the complex field $h$:
	\begin{equation}
		\begin{aligned}
		P_5(h,q)\equiv\det\mathcal H_5(q,h)=0,\\
		P_8(h,q)\equiv\det\mathcal H_8(q,h)=0.
		\label{eq:fibonacci_field_equations}
		\end{aligned}
	\end{equation}
	They are tenth- and sixteenth-degree equations in $h$. We solve them numerically using the BdG field equation in Eq.~\eqref{eq:field_pencil}, so each unit-cell momentum $q$ gives ten and sixteen field zeros, respectively. The real-field spectra give the phase diagrams, and the complex-field equations give the LYZ; both are obtained from the same Bloch BdG matrices.
	
	To understand why these numerical zeros repeatedly reach the real axis and reappear, consider two folded bands that are close to each other. Let $\epsilon_a$ and $\epsilon_b$ be their field roots without the anisotropic coupling, and let $g_{ab}$ be the coupling produced by the anisotropy. Near these two bands, the full field equation reduces to
	\begin{equation}
		\mathcal L_{ab}(h)=
		\begin{pmatrix}
			h-\epsilon_a&g_{ab}\\
			-g_{ab}^*&h-\epsilon_b
		\end{pmatrix},
		\label{eq:two_band}
	\end{equation}
	and $\det\mathcal L_{ab}(h)=0$ gives
	\begin{equation}
		h_\pm=\frac{\epsilon_a+\epsilon_b}{2}
		\pm\sqrt{\left(\frac{\epsilon_a-\epsilon_b}{2}\right)^2
			-|g_{ab}|^2}.
		\label{eq:two_band_roots}
	\end{equation}
	When
	\begin{equation}
		|\epsilon_a-\epsilon_b|>2|g_{ab}|,
	\end{equation}
	the two field roots are real. When
	\begin{equation}
		|\epsilon_a-\epsilon_b|<2|g_{ab}|,
	\end{equation}
	the square root is imaginary and the two real roots move away from the real axis as a complex-conjugate pair. The DM interaction changes the separation between the folded bands, while the anisotropy couples bands that are close to each other. As $D$ increases, two bands can first approach and then move apart, so complex zeros appear over a finite range of $D$. A Fibonacci cell contains more folded bands, allowing different pairs to meet at different values of $D$ and producing several such changes.
	
	For $\alpha=\gamma=0.5$, as used in Figs.~\ref{fig:fib5} and \ref{fig:fib8}, the numerical field equations give complex zeros in the following ranges within $0\leq D\leq1.25$:
	\begin{equation}
		\begin{aligned}
			F_l=5:\quad&
			0\leq D<0.1573,\\
			&0.2686<D<0.3044,\\
			&0.9554<D<1.2019;
		\end{aligned}
		\label{eq:fib5_complex_ranges}
	\end{equation}
	and
	\begin{equation}
		\begin{aligned}
			F_l=8:\quad&
			0\leq D<0.1547,\\
			&0.1975<D<0.2089,\\
			&0.3397<D<0.3826,\\
			&0.7433<D<0.9237.
		\end{aligned}
		\label{eq:fib8_complex_ranges}
	\end{equation}
	These separated ranges show directly that the complex zeros of the Fibonacci XY chains reach the real axis and reappear more than once. As the cell becomes longer, more pairs of folded bands can meet, and complex zeros disappear and reappear in several ranges of $D$.
	
	Figure~\ref{fig:fib5} shows the results for the $F_l=5$ Fibonacci XY chain. Figure~\ref{fig:fib5}(a) is the phase diagram obtained from the real-field spectrum, while Figs.~\ref{fig:fib5}(b) and \ref{fig:fib5}(c) show the long-distance correlation and vector chirality. The long-distance correlation identifies the antiferromagnetically ordered regions, while the gap and chirality together distinguish the paramagnetic and gapless chiral phases. The remaining panels show the LYZ as $D$ changes.
	
	At $D=0.10$, the zeros lie in the first complex range and form one large closed curve, as shown in Fig.~\ref{fig:fib5}(d). All the zeros become real at $D\simeq0.1573$. Complex zeros then reappear in the narrower range
	\begin{equation}
		0.2686<D<0.3044.
	\end{equation}
	Figure~\ref{fig:fib5}(e) uses $D=0.29$ and shows several smaller closed curves. They become real again above $D\simeq0.3044$. The value $D=0.60$ in Fig.~\ref{fig:fib5}(f) lies in a broad all-real range, where the zeros occupy several intervals of the real axis.
	
	When $D$ increases further to
	\begin{equation}
		0.9554<D<1.2019,
	\end{equation}
	another pair of folded bands approaches and complex zeros reappear. The value $D=1.08$ in Fig.~\ref{fig:fib5}(g) lies in this high-$D$ range, where the zeros form a large closed complex curve. Above $D\simeq1.2019$, this curve returns to the real axis, so all the zeros in Fig.~\ref{fig:fib5}(h) at $D=1.24$ are real. Every real-axis intersection of a complex curve and every endpoint of a real-zero interval agrees with a critical field in Fig.~\ref{fig:fib5}(a).
	
	Figure~\ref{fig:fib8} shows the results for the $F_l=8$ Fibonacci XY chain. This cell contains eight bonds, so the Bloch spectrum is divided into more bands. The real-field phase diagram in Fig.~\ref{fig:fib8}(a) contains more critical fields, while the long-distance correlation and vector chirality in Figs.~\ref{fig:fib8}(b) and \ref{fig:fib8}(c) identify the same three phases.
	
	At $D=0.10$, the zeros in Fig.~\ref{fig:fib8}(d) form one large closed curve. They reach the real axis at $D\simeq0.1547$ and then reappear briefly in the narrow range
	\begin{equation}
		0.1975<D<0.2089.
	\end{equation}
	As $D$ increases further, another set of complex zeros appears for
	\begin{equation}
		0.3397<D<0.3826.
	\end{equation}
	Figure~\ref{fig:fib8}(e) uses $D=0.36$ and shows several small closed curves produced by different folded bands. They return to the real axis above $D\simeq0.3826$. The value $D=0.60$ in Fig.~\ref{fig:fib8}(f) lies in an all-real range, with four separated real-zero intervals on the positive real axis.
	
	The larger high-$D$ complex range is
	\begin{equation}
		0.7433<D<0.9237.
	\end{equation}
	The value $D=0.83$ in Fig.~\ref{fig:fib8}(g) lies in this range, so a complex curve reappears. Above $D\simeq0.9237$, the zeros become real again, as shown at $D=1.00$ in Fig.~\ref{fig:fib8}(h). Compared with $F_l=5$, the $F_l=8$ cell has more narrow complex ranges, real-zero intervals, and critical fields. These changes come from the larger number of folded bands in the longer cell.
	
	Tong and Liu showed that varying the anisotropy $\gamma$ can split the LYZ of periodic and Fibonacci-modulated XY chains into several closed curves \cite{TongLiu2006}. Here $\gamma$ is fixed, and varying $D$ alone also moves the folded bands, causing the zero curves to split, reach the real axis, and reappear. The period-two XY chain has only two folded bands, and its zero curves become real after a single splitting. The period-three XY chain has an additional band meeting near $q\simeq\pi/2$, so complex zeros reappear once at larger $D$. The Fibonacci XY chains contain more folded bands, and different pairs meet at different values of $D$, allowing complex zeros to disappear and reappear several times. The two-band equation gives the common local explanation for these changes, while the complete phase boundaries and zero distributions are obtained from the corresponding Bloch BdG matrices.
	
	\section{Conclusions}
	\label{sec:conclusions}
	
	We employ the LYZ to investigate quantum phase transitions in XY spin-chain systems with the DM interaction, including the uniform chain, period-two and period-three modulated chains, as well as Fibonacci XY chains of lengths $F_l=5$ and $F_l=8$. The phase boundaries were obtained from the real-field spectrum, and the AFM, PM, and gapless chiral phases were identified using the excitation gap, long-distance correlation, and vector chirality. In every system, the real-axis intersections of the LYZ and the endpoints of the real-zero intervals coincide with the physical phase boundaries. The real-axis structure contains further information: isolated contacts give individual critical fields, whereas finite real-zero intervals coincide with gapless chiral regions. In the uniform XY chain, the DM interaction compresses an ellipse of zeros onto the real axis at $2D=\gamma$. In the period-two XY chain, increasing $D$ splits one closed curve into two before all the zeros become real, and the exact quartic equation gives both thresholds and the endpoints of the strong-DM gapless interval. In the period-three XY chain, a second meeting of folded bands near $q=\pi/2$ makes complex zeros reappear at larger $D$ after an all-real range. The Fibonacci XY chains contain more folded bands, and their complex zero curves disappear and reappear several times as $D$ increases, with more changes for the longer $F_l=8$ cell. The analytic field equations show that the DM interaction shifts the folded bands. When a particle band and a hole band approach each other, anisotropic pairing moves the corresponding two real field zeros away from the real axis as a complex-conjugate pair; when the two bands separate again, the zeros return to the real axis. These results demonstrate that LYZ can not only locate the quantum phase boundaries but also encode the intrinsic features of phase regions, enabling a complete description of the phase diagram for these modulated spin-chain systems. They are therefore of great value for detecting quantum phase transitions and characterizing quantum many-body states.
	
	\section*{Acknowledgments}
	We acknowledge financial support from National Natural Science Foundation of China (Grant Nos. 12074376 and 12505017), Beijing National Laboratory for Condensed Matter Physics (2025BNLCMPKF017), Beijing Municipal Natural Science Foundation (Grant No. 1222027), and the robotic AI-Scientist platform of Chinese Academy of Sciences.

	\appendix
	
	\section{Exact energy bands of the period-two XY chain}
	\label{app:p2_bands}
	
	The characteristic equation in Eq.~\eqref{eq:p2_energy_polynomial} is a fourth-order polynomial in $E$.  We write it as
	\begin{equation}
		E^4-C_2(q,h)E^2-L_2(q,h)E+P_2(h,q)=0,
		\label{eq:app_p2_compact_energy}
	\end{equation}
	where
	\begin{align}
		C_2(q,h)={}&2h^2+\frac12\Big[
		(1+\alpha^2)(1+\gamma^2)\nonumber\\
		&+2\alpha(1-\gamma^2)\cos q
		+8D^2(1-\cos q)\Big],
		\label{eq:app_p2_C}\\
		L_2(q,h)={}&4(1+\alpha)Dh\sin q,
		\label{eq:app_p2_L}\\
		P_2(h,q)={}&\det\mathcal H_2(q,h).
		\label{eq:app_p2_P}
	\end{align}
	The arguments $(q,h)$ are omitted below.  The term proportional to $E$ follows from the momentum asymmetry produced by the DM interaction.  Since $C_2$ and $P_2$ are even in $q$ while $L_2$ is odd, the energy roots satisfy $\{E_n(-q)\}=\{-E_n(q)\}$.  At a general momentum, the polynomial can be factored as
	\begin{align}
		E^4-C_2E^2-L_2E+P_2
		={}&(E^2+sE+t)\nonumber\\
		&\times(E^2-sE+u).
		\label{eq:app_p2_factorization}
	\end{align}
	Matching the coefficients of equal powers of $E$ gives
	\begin{equation}
		t+u-s^2=-C_2,\qquad
		s(u-t)=-L_2,\qquad
		tu=P_2.
		\label{eq:app_p2_matching}
	\end{equation}
	Setting $z=s^2$, the first two relations give
	\begin{equation}
		t=\frac12\left(z-C_2+\frac{L_2}{\sqrt z}\right),
		\qquad
		u=\frac12\left(z-C_2-\frac{L_2}{\sqrt z}\right).
		\label{eq:app_p2_tu}
	\end{equation}
	Substitution into $tu=P_2$ leads to the cubic equation
	\begin{equation}
		z^3-2C_2z^2+
		\left(C_2^2-4P_2\right)z-L_2^2=0.
		\label{eq:app_p2_resolvent}
	\end{equation}
	For any nonzero solution $z$ of Eq.~\eqref{eq:app_p2_resolvent}, Eq.~\eqref{eq:app_p2_compact_energy} becomes
	\begin{align}
		{}&
		\left[
		E^2+\sqrt z\,E+
		\frac12\left(z-C_2+\frac{L_2}{\sqrt z}\right)
		\right]\nonumber\\
		&\times
		\left[
		E^2-\sqrt z\,E+
		\frac12\left(z-C_2-\frac{L_2}{\sqrt z}\right)
		\right]=0.
		\label{eq:app_p2_exact_factors}
	\end{align}
	The four energy bands are therefore
	\begin{align}
		E_1(q,h)&=
		\frac12\left[
		-\sqrt z+
		\sqrt{2C_2-z-\frac{2L_2}{\sqrt z}}
		\right],
		\nonumber\\
		E_2(q,h)&=
		\frac12\left[
		-\sqrt z-
		\sqrt{2C_2-z-\frac{2L_2}{\sqrt z}}
		\right],
		\nonumber\\
		E_3(q,h)&=
		\frac12\left[
		\sqrt z+
		\sqrt{2C_2-z+\frac{2L_2}{\sqrt z}}
		\right],
		\nonumber\\
		E_4(q,h)&=
		\frac12\left[
		\sqrt z-
		\sqrt{2C_2-z+\frac{2L_2}{\sqrt z}}
		\right].
		\label{eq:app_p2_energy_roots}
	\end{align}
	
	The cubic equation for $z$ can also be solved in radicals.  Define
	\begin{align}
		\Omega_2&=C_2^2+12P_2,
		\label{eq:app_p2_Omega}\\
		\Lambda_2&=-2C_2^3+72C_2P_2+27L_2^2,
		\label{eq:app_p2_Lambda}\\
		\mathcal U_2&=
		\left[
		\frac{\Lambda_2+
			\sqrt{\Lambda_2^2-4\Omega_2^3}}{2}
		\right]^{1/3}.
		\label{eq:app_p2_U}
	\end{align}
	Choosing a cube-root branch with $\mathcal U_2\neq0$, one solution of Eq.~\eqref{eq:app_p2_resolvent} is
	\begin{equation}
		z=\frac13\left(
		2C_2+\mathcal U_2+\frac{\Omega_2}{\mathcal U_2}
		\right).
		\label{eq:app_p2_z_radical}
	\end{equation}
	The three cube-root branches give the three solutions of the cubic equation.  At a degenerate point with $\mathcal U_2=0$, $z$ follows directly from Eq.~\eqref{eq:app_p2_resolvent}.
	
	Whenever $L_2=0$, the energy equation is quadratic in $E^2$.  This occurs at $q=0$ and $\pi$, and also for all $q$ when $D=0$.  The four bands then simplify to
	\begin{equation}
		E=\pm\sqrt{
			\frac{C_2\pm\sqrt{C_2^2-4P_2}}{2}}.
		\label{eq:app_p2_symmetric_roots}
	\end{equation}
	The two signs are chosen independently.  For real $h$, these expressions give the four real eigenvalues of the Hermitian BdG matrix.  Setting $E=0$ in Eq.~\eqref{eq:app_p2_compact_energy} gives $P_2(h,q)=0$, which is the field equation used for the LYZ in Sec.~\ref{sec:period2}.
	
	\section{Energy bands and high-\texorpdfstring{$D$}{D} roots of the period-three XY chain}
	\label{app:p3}
	
	The characteristic equation of the $6\times6$ BdG matrix in Eqs.~\eqref{eq:p3_normal_block} and \eqref{eq:p3_pairing_block} can be written as
	\begin{equation}
		E^6-C_3E^4-L_3E^3+M_3E^2+N_3E+P_3=0,
		\label{eq:app_p3_energy}
	\end{equation}
	where
	\begin{align}
		C_3(h)={}&3h^2+\frac12(1+2\alpha^2)(1+\gamma^2)+6D^2,
		\label{eq:app_p3_C}\\
		L_3(q)={}&D\sin q
		\left[4D^2+\alpha(\alpha+2)(\gamma^2-1)\right],
		\label{eq:app_p3_L}
	\end{align}
	\begin{align}
		M_3(q,h)={}&3h^4+(1+2\alpha^2)\gamma^2h^2\nonumber\\
		&-\frac32h\cos q
		\left[\alpha^2(\gamma^2-1)+4(1+2\alpha)D^2\right]\nonumber\\
		&+\frac{(1+2\alpha^2)^2(1+\gamma^4)}{16}\nonumber\\
		&+\frac{\gamma^2(4\alpha^4+12\alpha^2-1)}8
		+\frac32D^2(1+2\alpha^2)\nonumber\\
		&+9D^4
		+\frac12D^2\gamma^2(1-8\alpha-2\alpha^2).
		\label{eq:app_p3_M}
	\end{align}
	\begin{multline}
		N_3(q,h)=D\sin q\Bigg\{
		\Big[\alpha(\alpha+2)(\gamma^2+3)-12D^2\Big]h^2\\
		+\frac{\alpha(\gamma^2-1)}{4}
		\Big[2(\alpha^3+1)(\gamma^2+1)
		+4\alpha^2+\alpha(1-\gamma^2)\Big]\\
		+12D^4-D^2\Big[
		\alpha^2(3\gamma^2+1)+2\alpha(\gamma^2+3)
		+\gamma^2-1\Big]\Bigg\}.
		\label{eq:app_p3_N}
	\end{multline}
	\begin{align}
		P_3(h,q)={}&-\frac1{16}
		\left[4h^3+A_3h+B_3\cos q\right]^2\nonumber\\
		&-\frac1{16}
		F_\alpha(D,\gamma)F_\alpha(-D,\gamma)\sin^2q .
		\label{eq:app_p3_P}
	\end{align}
	Here $A_3$, $B_3$, and $F_\alpha(D,\gamma)$ are given in Eqs.~\eqref{eq:p3_A}, \eqref{eq:p3_B}, and \eqref{eq:p3_F}.  For general $q$ and parameters, Eq.~\eqref{eq:app_p3_energy} cannot be solved in radicals, so its six energy roots are obtained by numerically diagonalizing $\mathcal H_3(q,h)$.
	
	At $q=0$ and $\pi$, $\sin q=0$ and hence $L_3=N_3=0$.  Setting $z=E^2$ reduces Eq.~\eqref{eq:app_p3_energy} to
	\begin{equation}
		z^3-C_3z^2+M_3z+P_3=0,
		\qquad q=0,\pi .
		\label{eq:app_p3_z_cubic}
	\end{equation}
	The same reduction holds at every momentum when $D=0$.  With $\omega=e^{2\pi\ii/3}$, define
	\begin{align}
		\Lambda_3(q,h)&=2C_3^3-9C_3M_3-27P_3,
		\label{eq:app_p3_Lambda}\\
		\mathcal U_3(q,h)&=
		\left[
		\frac{\Lambda_3+
			\sqrt{\Lambda_3^2-4(C_3^2-3M_3)^3}}{2}
		\right]^{1/3}.
		\label{eq:app_p3_U}
	\end{align}
	For $\nu=0,1,2$, the six energy bands at $q=0$ and $\pi$ are
	\begin{equation}
		E_\nu(q,h)=\pm\frac1{\sqrt3}\left[
		C_3+\omega^\nu\mathcal U_3
		+\frac{C_3^2-3M_3}{\omega^\nu\mathcal U_3}
		\right]^{1/2}.
		\label{eq:app_p3_energy_roots}
	\end{equation}
	The three values of $\nu$ give the three roots for $E^2$, and the two signs give the six energy bands.  At a degenerate point with $\mathcal U_3=0$, the roots follow directly from Eq.~\eqref{eq:app_p3_z_cubic}.
	
	Setting $E=0$ in Eq.~\eqref{eq:app_p3_energy} gives the field equation $P_3(h,q)=0$.  At $q=0$ and $\pi$, this gives Eq.~\eqref{eq:p3_contact_cubic}, whose cubic discriminant is
	\begin{equation}
		\mathcal D_3=-\frac{A_3^3+27B_3^2}{16}.
		\label{eq:app_p3_discriminant}
	\end{equation}
	Its zeros mark changes in the number of real contacts at $q=0$ or $\pi$.
	
	To obtain the high-$D$ complex roots near $q=\pi/2$, write $D=D_*+\delta D$ and expand Eq.~\eqref{eq:p3_resonance} for small $\delta D$, $\gamma$, and $h$.  For a general $\alpha$,
	\begin{align}
		h^2={}&
		\frac{16\alpha^2(\alpha+2)^2}
		{(\alpha+1)^2(5\alpha+1)^2}\,\delta D^2\nonumber\\
		&-\frac{4\alpha^2(1-\alpha)^2}
		{(5\alpha+1)^2}\,\gamma^2+\cdots .
		\label{eq:app_p3_expansion}
	\end{align}
	The roots are complex in the interval
	\begin{equation}
		|D-D_*|<
		\frac{|1-\alpha^2|}{2(\alpha+2)}|\gamma|,
		\label{eq:app_p3_width}
	\end{equation}
	and their imaginary part at the center is
	\begin{equation}
		|\Imm h(D_*)|=
		\frac{2\alpha|1-\alpha|}{5\alpha+1}|\gamma|.
		\label{eq:app_p3_height}
	\end{equation}
	Both quantities vanish at $\alpha=1$, where the period-three cell is only a folded representation of the uniform XY chain.  Exchange modulation is therefore essential for the second complex curve.
	
	\bibliography{references}

@article{YangLee1952,
  author = {Yang, C. N. and Lee, T. D.},
  title = {Statistical Theory of Equations of State and Phase Transitions. {I}. {Theory} of Condensation},
  journal = {Phys. Rev.},
  volume = {87},
  pages = {404--409},
  year = {1952},
  doi = {10.1103/PhysRev.87.404}
}

@article{LeeYang1952,
  author = {Lee, T. D. and Yang, C. N.},
  title = {Statistical Theory of Equations of State and Phase Transitions. {II}. {Lattice} Gas and {Ising} Model},
  journal = {Phys. Rev.},
  volume = {87},
  pages = {410--419},
  year = {1952},
  doi = {10.1103/PhysRev.87.410}
}

@article{ItzyksonPearsonZuber1983,
  author = {Itzykson, Claude and Pearson, Robert B. and Zuber, Jean-Bernard},
  title = {Distribution of Zeros in {Ising} and Gauge Models},
  journal = {Nucl. Phys. B},
  volume = {220},
  pages = {415--433},
  year = {1983},
  doi = {10.1016/0550-3213(83)90499-6}
}

@article{SuzukiFisher1971,
  author = {Suzuki, Masuo and Fisher, Michael E.},
  title = {Zeros of the Partition Function for the {Heisenberg}, Ferroelectric, and General {Ising} Models},
  journal = {J. Math. Phys.},
  volume = {12},
  pages = {235--246},
  year = {1971},
  doi = {10.1063/1.1665583}
}

@article{TongLiu2006,
  author = {Tong, Peiqing and Liu, Xiaoxian},
  title = {{Lee--Yang} Zeros of Periodic and Quasiperiodic Anisotropic {XY} Chains in a Transverse Field},
  journal = {Phys. Rev. Lett.},
  volume = {97},
  pages = {017201},
  year = {2006},
  doi = {10.1103/PhysRevLett.97.017201}
}

@article{ZhongLiuTong2007,
  author = {Zhong, Ming and Liu, Xiaoxian and Tong, Peiqing},
  title = {{Lee--Yang} Zeros and Quantum Phase Transitions of Nonuniform Anisotropic Chains in a Transverse Field},
  journal = {Int. J. Mod. Phys. B},
  volume = {21},
  pages = {4225--4229},
  year = {2007},
  doi = {10.1142/S021797920704544X}
}

@article{TongZhong2001,
  author = {Tong, Peiqing and Zhong, Ming},
  title = {Quantum Phase Transitions of Periodic Anisotropic {XY} Chain in a Transverse Field},
  journal = {Physica B},
  volume = {304},
  pages = {91--106},
  year = {2001},
  doi = {10.1016/S0921-4526(01)00546-4}
}

@article{TongZhong2002,
  author = {Tong, Peiqing and Zhong, Ming},
  title = {Quantum Phase Transitions of a Quasiperiodic Anisotropic {XY} Chain in a Transverse Magnetic Field},
  journal = {Phys. Rev. B},
  volume = {65},
  pages = {064421},
  year = {2002},
  doi = {10.1103/PhysRevB.65.064421}
}

@article{SatijaDoria1988,
  author = {Satija, I. I. and Doria, M. M.},
  title = {Quasiperiodic Anisotropic {XY} Model},
  journal = {Phys. Rev. B},
  volume = {38},
  pages = {5174--5176},
  year = {1988},
  doi = {10.1103/PhysRevB.38.5174}
}

@article{deLima2007,
  author = {de Lima, J. P. and Gon{\c{c}}alves, L. L. and Alves, T. F. A.},
  title = {Anisotropic {XY} Model on the Inhomogeneous Periodic Chain},
  journal = {Phys. Rev. B},
  volume = {75},
  pages = {214406},
  year = {2007},
  doi = {10.1103/PhysRevB.75.214406}
}

@article{Liu2011,
  author = {Liu, Ben-Qiong and Shao, Bin and Li, Jun-Gang and Zou, Jian and Wu, Lian-Ao},
  title = {Quantum and Classical Correlations in the One-Dimensional {XY} Model with {Dzyaloshinskii--Moriya} Interaction},
  journal = {Phys. Rev. A},
  volume = {83},
  pages = {052112},
  year = {2011},
  doi = {10.1103/PhysRevA.83.052112}
}

@article{Zhong2013,
  author = {Zhong, Ming and Xu, Hui and Liu, Xiao-Xian and Tong, Pei-Qing},
  title = {The Effects of the {Dzyaloshinskii--Moriya} Interaction on the Ground-State Properties of the {XY} Chain in a Transverse Field},
  journal = {Chin. Phys. B},
  volume = {22},
  pages = {090313},
  year = {2013},
  doi = {10.1088/1674-1056/22/9/090313}
}

@article{Peng2015,
  author = {Peng, Xinhua and Zhou, Hui and Wei, Bo-Bo and Cui, Jiangyu and Du, Jiangfeng and Liu, Ren-Bao},
  title = {Experimental Observation of {Lee--Yang} Zeros},
  journal = {Phys. Rev. Lett.},
  volume = {114},
  pages = {010601},
  year = {2015},
  doi = {10.1103/PhysRevLett.114.010601}
}

@article{Francis2021,
  author = {Francis, Akhil and Zhu, Daiwei and Huerta Alderete, C. and Johri, Sonika and Xiao, X. and Freericks, J. K. and Monroe, Christopher and Linke, N. M. and Kemper, A. F.},
  title = {Many-Body Thermodynamics on Quantum Computers via Partition Function Zeros},
  journal = {Sci. Adv.},
  volume = {7},
  pages = {eabf2447},
  year = {2021},
  doi = {10.1126/sciadv.abf2447}
}

@article{Kist2021,
  author = {Kist, Timo and Lado, Jose L. and Flindt, Christian},
  title = {{Lee--Yang} Theory of Criticality in Interacting Quantum Many-Body Systems},
  journal = {Phys. Rev. Research},
  volume = {3},
  pages = {033206},
  year = {2021},
  doi = {10.1103/PhysRevResearch.3.033206}
}

@article{VecseiLadoFlindt2022,
  author = {Vecsei, Pascal M. and Lado, Jose L. and Flindt, Christian},
  title = {{Lee--Yang} Theory of the Two-Dimensional Quantum {Ising} Model},
  journal = {Phys. Rev. B},
  volume = {106},
  pages = {054402},
  year = {2022},
  doi = {10.1103/PhysRevB.106.054402}
}

@article{Li2025,
  author = {Li, Hongchao},
  title = {{Yang--Lee} Zeros in Quantum Phase Transitions: An Entanglement Perspective},
  journal = {Phys. Rev. B},
  volume = {111},
  pages = {045139},
  year = {2025},
  doi = {10.1103/PhysRevB.111.045139}
}

@article{GuSun2026,
  author = {Gu, Tian-Yi and Sun, Gaoyong},
  title = {Fidelity Zeros and {Lee--Yang} Theory of Quantum Phase Transitions},
  journal = {Phys. Rev. B},
  volume = {113},
  pages = {014417},
  year = {2026},
  doi = {10.1103/3x34-f53v}
}

@article{BenaDrozLipowski2005,
  author  = {Bena, Ioana and Droz, Michel and Lipowski, Adam},
  title   = {Statistical Mechanics of Equilibrium and Nonequilibrium Phase Transitions: The {Yang--Lee} Formalism},
  journal = {Int. J. Mod. Phys. B},
  volume  = {19},
  pages   = {4269--4329},
  year    = {2005},
  doi     = {10.1142/S0217979205032759}
}

@article{WeiLiu2012,
  author  = {Wei, Bo-Bo and Liu, Ren-Bao},
  title   = {{Lee--Yang} Zeros and Critical Times in Decoherence of a Probe Spin Coupled to a Bath},
  journal = {Phys. Rev. Lett.},
  volume  = {109},
  pages   = {185701},
  year    = {2012},
  doi     = {10.1103/PhysRevLett.109.185701}
}

@article{Dzyaloshinsky1958,
  author  = {Dzyaloshinsky, I.},
  title   = {A Thermodynamic Theory of ``Weak'' Ferromagnetism of Antiferromagnetics},
  journal = {J. Phys. Chem. Solids},
  volume  = {4},
  pages   = {241--255},
  year    = {1958},
  doi     = {10.1016/0022-3697(58)90076-3}
}

@article{Moriya1960,
  author  = {Moriya, T.},
  title   = {Anisotropic Superexchange Interaction and Weak Ferromagnetism},
  journal = {Phys. Rev.},
  volume  = {120},
  pages   = {91--98},
  year    = {1960},
  doi     = {10.1103/PhysRev.120.91}
}

@article{LiebSchultzMattis1961,
  author = {Lieb, Elliott and Schultz, Theodore and Mattis, Daniel},
  title = {Two Soluble Models of an Antiferromagnetic Chain},
  journal = {Ann. Phys. (N.Y.)},
  volume = {16},
  pages = {407--466},
  year = {1961},
  doi = {10.1016/0003-4916(61)90115-4}
}

@article{BarouchMcCoy1971,
  author = {Barouch, Eytan and McCoy, Barry M.},
  title = {Statistical Mechanics of the {XY} Model. {II}. {Spin}-Correlation Functions},
  journal = {Phys. Rev. A},
  volume = {3},
  pages = {786--804},
  year = {1971},
  doi = {10.1103/PhysRevA.3.786}
}

@article{Luck1993,
  author = {Luck, Jean-Marc},
  title = {Critical Behavior of the Aperiodic Quantum {Ising} Chain in a Transverse Magnetic Field},
  journal = {J. Stat. Phys.},
  volume = {72},
  pages = {417--458},
  year = {1993},
  doi = {10.1007/BF01048019}
}

@article{Hermisson2000,
  author = {Hermisson, Joachim},
  title = {Aperiodic and Correlated Disorder in {XY} Chains: Exact Results},
  journal = {J. Phys. A: Math. Gen.},
  volume = {33},
  pages = {57--79},
  year = {2000},
  doi = {10.1088/0305-4470/33/1/304}
}

@article{BaakeGrimmPisani1995,
  author = {Baake, Michael and Grimm, Uwe and Pisani, Carmelo},
  title = {Partition Function Zeros for Aperiodic Systems},
  journal = {J. Stat. Phys.},
  volume = {78},
  pages = {285--297},
  year = {1995},
  doi = {10.1007/BF02183349}
}

@article{BarataGoldbaum2001,
  author = {Barata, Jo{\~a}o Carlos Alves and Goldbaum, Pedro Silva},
  title = {On the Distribution and Gap Structure of {Lee--Yang} Zeros for the {Ising} Model: Periodic and Aperiodic Couplings},
  journal = {J. Stat. Phys.},
  volume = {103},
  pages = {857--891},
  year = {2001},
  doi = {10.1023/A:1010332500031}
}

@article{DerzhkoVerkholyakKrokhmalskiiButtner2006,
  author = {Derzhko, Oleg and Verkholyak, Taras and Krokhmalskii, Taras and B{\"u}ttner, Helmut},
  title = {Dynamic Probes of Quantum Spin Chains with the {Dzyaloshinskii--Moriya} Interaction},
  journal = {Phys. Rev. B},
  volume = {73},
  pages = {214407},
  year = {2006},
  doi = {10.1103/PhysRevB.73.214407}
}

@article{JafariKargarianLangariSiahatgar2008,
  author = {Jafari, R. and Kargarian, M. and Langari, A. and Siahatgar, M.},
  title = {Phase Diagram and Entanglement of the {Ising} Model with {Dzyaloshinskii--Moriya} Interaction},
  journal = {Phys. Rev. B},
  volume = {78},
  pages = {214414},
  year = {2008},
  doi = {10.1103/PhysRevB.78.214414}
}

@article{KadarZimboras2010,
  author = {K{\'a}d{\'a}r, Zolt{\'a}n and Zimbor{\'a}s, Zolt{\'a}n},
  title = {Entanglement Entropy in Quantum Spin Chains with Broken Reflection Symmetry},
  journal = {Phys. Rev. A},
  volume = {82},
  pages = {032334},
  year = {2010},
  doi = {10.1103/PhysRevA.82.032334}
}

@article{MatsumotoNakagawaUeda2022,
  author = {Matsumoto, Norifumi and Nakagawa, Masaya and Ueda, Masahito},
  title = {Embedding the {Yang--Lee} Quantum Criticality in Open Quantum Systems},
  journal = {Phys. Rev. Research},
  volume = {4},
  pages = {033250},
  year = {2022},
  doi = {10.1103/PhysRevResearch.4.033250}
}

@article{VecseiFlindtLado2023,
  author = {Vecsei, Pascal M. and Flindt, Christian and Lado, Jose L.},
  title = {{Lee--Yang} Theory of Quantum Phase Transitions with Neural Network Quantum States},
  journal = {Phys. Rev. Research},
  volume = {5},
  pages = {033116},
  year = {2023},
  doi = {10.1103/PhysRevResearch.5.033116}
}

@article{TaoEtAl2023,
  author = {Tao, Hong and Su, Yuguo and Zhang, Xingyu and Liu, Jing and Wang, Xiaoguang},
  title = {{Lee--Yang} Zeros and Quantum {Fisher} Information Matrix in a Nonlinear System},
  journal = {Phys. Rev. E},
  volume = {108},
  pages = {024104},
  year = {2023},
  doi = {10.1103/PhysRevE.108.024104}
}

@article{GaoEtAl2024,
  author = {Gao, Huixia and Wang, Kunkun and Xiao, Lei and Nakagawa, Masaya and Matsumoto, Norifumi and Qu, Dengke and Lin, Haiqing and Ueda, Masahito and Xue, Peng},
  title = {Experimental Observation of the {Yang--Lee} Quantum Criticality in Open Quantum Systems},
  journal = {Phys. Rev. Lett.},
  volume = {132},
  pages = {176601},
  year = {2024},
  doi = {10.1103/PhysRevLett.132.176601}
}

@article{LanEtAl2024,
  author = {Lan, Ziheng and Liu, Wenquan and Wu, Yang and Ye, Xiangyu and Yang, Zhesen and Duan, Chang-Kui and Wang, Ya and Rong, Xing},
  title = {Experimental Investigation of {Lee--Yang} Criticality Using a {non-Hermitian} Quantum System},
  journal = {Chin. Phys. Lett.},
  volume = {41},
  pages = {050301},
  year = {2024},
  doi = {10.1088/0256-307X/41/5/050301}
}

@article{MahdavifarEtAl2024,
  author = {Mahdavifar, Saeed and Salehpour, Mahboubeh and Cheraghi, Hadi and Afrousheh, Kourosh},
  title = {Resilience of Quantum Spin Fluctuations against {Dzyaloshinskii--Moriya} Interaction},
  journal = {Sci. Rep.},
  volume = {14},
  pages = {10034},
  year = {2024},
  doi = {10.1038/s41598-024-60502-y}
}

@article{VecseiLadoFlindt2025,
  author = {Vecsei, Pascal M. and Lado, Jose L. and Flindt, Christian},
  title = {{Lee--Yang} Formalism for Phase Transitions of Interacting Fermions Using Tensor Networks},
  journal = {Phys. Rev. B},
  volume = {111},
  pages = {075134},
  year = {2025},
  doi = {10.1103/PhysRevB.111.075134}
}

@article{LvEtAl2026,
  author = {Lv, Songtai and Liu, Yang and Zhao, Erhai and Zou, Haiyuan and Xiang, Tao},
  title = {Giant Bubbles of {Fisher} Zeros in the Quantum {XY} Chain},
  journal = {Phys. Rev. B},
  volume = {113},
  pages = {L241117},
  year = {2026},
  doi = {10.1103/pwcc-75fn}
}

@article{LiuLvMengTanZhaoZou2024,
  author = {Liu, Yang and Lv, Songtai and Meng, Yuchen and Tan, Zefan and Zhao, Erhai and Zou, Haiyuan},
  title = {Exact {Fisher} Zeros and Thermofield Dynamics across a Quantum Critical Point},
  journal = {Phys. Rev. Research},
  volume = {6},
  pages = {043139},
  year = {2024},
  doi = {10.1103/PhysRevResearch.6.043139}
}

@article{WadaKitazawaKanaya2025,
  author = {Wada, Tatsuya and Kitazawa, Masakiyo and Kanaya, Kazuyuki},
  title = {Locating Critical Points Using Ratios of {Lee--Yang} Zeros},
  journal = {Phys. Rev. Lett.},
  volume = {134},
  pages = {162302},
  year = {2025},
  doi = {10.1103/PhysRevLett.134.162302}
}

@article{ZhangMaoHuZhaoSunYou2025,
  author = {Zhang, Wen-Yi and Mao, Meng-Yun and Hu, Qing-Min and Zhao, Xinzhi and Sun, Gaoyong and You, Wen-Long},
  title = {{Yang--Lee} Edge Singularity and Quantum Criticality in a {non-Hermitian} {PXP} Model},
  journal = {Phys. Rev. B},
  volume = {112},
  pages = {155135},
  year = {2025},
  doi = {10.1103/vlfm-jfq5}
}

@article{YiDingRenWangYou2018,
  author = {Yi, Tian-Cheng and Ding, Yue-Ran and Ren, Jie and Wang, Yi-Min and You, Wen-Long},
  title = {Quantum Coherence of the {XY} Model with {Dzyaloshinskii--Moriya} Interaction},
  journal = {Acta Phys. Sin.},
  volume = {67},
  pages = {140303},
  year = {2018},
  doi = {10.7498/aps.67.20172755}
}

@article{WangYanYi2010,
  author = {Wang, Lin-Cheng and Yan, Jun-Yan and Yi, Xue-Xi},
  title = {Geometric Phases and Quantum Phase Transitions in Inhomogeneous {XY} Spin Chains: Effect of the {Dzyaloshinsky--Moriya} Interaction},
  journal = {Chin. Phys. B},
  volume = {19},
  pages = {040512},
  year = {2010},
  doi = {10.1088/1674-1056/19/4/040512}
}

@article{ChatterjeeEtAl2024,
  author = {Chatterjee, Arijit and Mahesh, T. S. and Nisse, Mounir and Lim, Yen-Kheng},
  title = {Observing Algebraic Variety of {Lee--Yang} Zeros in Asymmetrical Systems via a Quantum Probe},
  journal = {Phys. Rev. A},
  volume = {109},
  pages = {062601},
  year = {2024},
  doi = {10.1103/PhysRevA.109.062601}
}

@article{RoyEtAl2019,
  author = {Roy, Saptarshi and Chanda, Titas and Das, Tamoghna and Sadhukhan, Debasis and Sen(De), Aditi and Sen, Ujjwal},
  title = {Phase Boundaries in an Alternating-Field Quantum {XY} Model with {Dzyaloshinskii--Moriya} Interaction: Sustainable Entanglement in Dynamics},
  journal = {Phys. Rev. B},
  volume = {99},
  pages = {064422},
  year = {2019},
  doi = {10.1103/PhysRevB.99.064422}
}

@article{CornagliaEtAl2024,
  author = {Cornaglia, Pablo S. and Chinellato, Leandro M. and Batista, Cristian D.},
  title = {Quasicrystalline Chiral Soliton Lattices in a {Fibonacci} Helimagnet},
  journal = {Phys. Rev. B},
  volume = {110},
  pages = {054430},
  year = {2024},
  doi = {10.1103/PhysRevB.110.054430}
}

@article{TaoEtAl2022,
  author = {Tao, Hong and Shao, Lei and Zhang, Zhucheng and Zhang, Xingyu and Zhang, Rui and Chen, Jie and Lu, Wangjun},
  title = {Quantum {Fisher} Information and {Lee--Yang} Zeros with Quantum Coherence of the Isotropic {XY} Model and the Energy Scale},
  journal = {Ann. Phys. (Berlin)},
  volume = {534},
  pages = {2200291},
  year = {2022},
  doi = {10.1002/andp.202200291}
}

@article{LiuLvYangZou2023,
  author = {Liu, Yang and Lv, Songtai and Yang, Yang and Zou, Haiyuan},
  title = {Signatures of Quantum Criticality in the Complex Inverse Temperature Plane},
  journal = {Chin. Phys. Lett.},
  volume = {40},
  pages = {050502},
  year = {2023},
  doi = {10.1088/0256-307X/40/5/050502}
}

@article{LiYang2023,
  author = {Li, Chengshu and Yang, Fan},
  title = {{Lee--Yang} Zeros in the {Rydberg} Atoms},
  journal = {Front. Phys.},
  volume = {18},
  pages = {22301},
  year = {2023},
  doi = {10.1007/s11467-022-1226-6}
}

@article{ShenEtAl2023,
  author = {Shen, Ruizhe and Chen, Tianqi and Aliyu, Mohammad Mujahid and Qin, Fang and Zhong, Yin and Loh, Huanqian and Lee, Ching Hua},
  title = {Proposal for Observing {Yang--Lee} Criticality in {Rydberg} Atomic Arrays},
  journal = {Phys. Rev. Lett.},
  volume = {131},
  pages = {080403},
  year = {2023},
  doi = {10.1103/PhysRevLett.131.080403}
}

@article{TimoninChitov2021,
  author = {Timonin, P. N. and Chitov, Gennady Y.},
  title = {Disorder Lines, Modulation, and Partition Function Zeros in Free Fermion Models},
  journal = {Phys. Rev. B},
  volume = {104},
  pages = {045106},
  year = {2021},
  doi = {10.1103/PhysRevB.104.045106}
}

@article{ChitovGadgeTimonin2022,
  author = {Chitov, Gennady Y. and Gadge, Karun and Timonin, P. N.},
  title = {Disentanglement, Disorder Lines, and {Majorana} Edge States in a Solvable Quantum Chain},
  journal = {Phys. Rev. B},
  volume = {106},
  pages = {125146},
  year = {2022},
  doi = {10.1103/PhysRevB.106.125146}
}

@article{CaoFuLiuZhongTong2024,
  author = {Cao, Kaiyuan and Fu, Hao and Liu, Xue and Zhong, Ming and Tong, Peiqing},
  title = {Quantum Phase Transitions in the Alternating {XY} Chain with Three-Site Interactions},
  journal = {Physica B},
  volume = {682},
  pages = {415876},
  year = {2024},
  doi = {10.1016/j.physb.2024.415876}
}

@article{LiuYinChen2024,
  author = {Liu, Jinghu and Yin, Shuai and Chen, Li},
  title = {Imaginary-Temperature Zeros for Quantum Phase Transitions},
  journal = {Phys. Rev. B},
  volume = {110},
  pages = {134313},
  year = {2024},
  doi = {10.1103/PhysRevB.110.134313}
}
	
\end{document}